\documentclass[aps,pra,showpacs,superscriptaddress,preprint]{revtex4}
\usepackage{amsmath}
\usepackage{graphicx}

\catcode`ð=\active
 \defð{\u{g}}
 \catcode`Ð=\active
 \defÐ{\u{G}}
 \catcode`Ý=\active
\defÝ{\. I}
 \catcode`ö=\active
\defö{\"{o}}
 \catcode`Ö=\active
 \defÖ{\"O}
 \catcode`ü=\active
 \defü{\"{u}}
 \catcode`Ü=\active
 \defÜ{\"{U}}
 \catcode`Þ=\active
\defÞ{\c{S}}
 \catcode`þ=\active
 \defþ{\c{s}}
 \catcode`ý=\active
 \defý{{\i}}
 \catcode`ç=\active
\defç{\d{c}}
 \catcode`Ç=\active
\defÇ{\d{C}}

\begin{document}

\title{I\textbf{mproved analytical approximation to arbitrary }$l$\textbf{%
-state solutions of the Schr\"{o}dinger equation for the hyperbolical
potentials }}
\author{{\small Sameer M. Ikhdair}}
\email[E-mail: ]{sikhdair@neu.edu.tr}
\affiliation{Department of Physics, Near East University, Nicosia, North Cyprus, Turkey}
\author{Ramazan Sever}
\email[E-mail: ]{sever@metu.edu.tr}
\affiliation{Department of Physics, Middle East Technical University, 06800 Ankara,Turkey}
\date{%
%TCIMACRO{\TeXButton{today}{\today}}%
%BeginExpansion
\today%
%EndExpansion
}

\begin{abstract}
A new approximation scheme to the centrifugal term is proposed to obtain the
$l\neq 0$ bound-state solutions of the Schr\"{o}dinger equation for an
exponential-type potential in the framework of the hypergeometric method.
The corresponding normalized wave functions are also found in terms of the
Jacobi polynomials. To show the accuracy of the new proposed approximation
scheme, we calculate the energy eigenvalues numerically for arbitrary
quantum numbers $n$ and $l$ with two different values of the potential
parameter $\sigma _{\text{0}}.$ Our numerical results are of high accuracy
like the other numerical results obtained by using program based on a
numerical integration procedure for short-range and long-range potentials.
The energy bound-state solutions for the $s$-wave ($l=0$) and $\sigma _{0}=1$
cases are given.\newline
\end{abstract}

\maketitle

\newpage

\section{Introduction}

The exact analytic solutions of the wave equations (nonrelativistic and
relativistic) are only possible for certain potentials of physical interest
under consideration since they contain all the necessary information on the
quantum system. It is well known that the exact solutions of these wave
equations are only possible in a few simple cases such as the Coulomb, the
harmonic oscillator, pseudoharmonic potentials and others [1-5]. Recently,
the analytic exact solutions of the wave equation with some exponential-type
potentials are impossible for $l\neq 0$ states. Approximation methods have
to be used to deal with the centrifugal term like the Pekeris approximation
[6-8] and the approximated scheme suggested by Greene and Aldrich [9]. Some
of these exponential-type potentials include the Morse potential [10], the
Hulth\'{e}n potential [11], the P\"{o}schl-Teller [12], the Woods-Saxon
potential [13], the Kratzer-type and pseudoharmonic potentials [14], the
Rosen-Morse-type potentials [15], the Manning-Rosen potential [15-22], other
multiparameter exponential-type potentials [23,24] and hyperbolical
potential [25-27].

In this work, we attempt to study another exponential-type potential called
the hyperbolical potential [25-27]
\begin{equation}
V(r)=D\left[ 1-\sigma _{0}\coth (\alpha r)\right] ,
\end{equation}%
where $D,$ $\alpha $ and $\sigma _{0}$ are three positive parameters. It is
indicated in [25] that this exponential-type potential is closely related to
the Morse, the Kratzer, the Coulomb, the harmonic oscillator and other
potential functions. The properties and applications of this potential are
given in [25,26]. It is known that for this potential the Schr\"{o}dinger
equation (SE) can be solved for the $s$-wave, angular momentum quantum
number $l=0.$ However, for a general solution, it is need to include some
approximations if one wants to obtain analytical or semianalytical solutions
to the SE. For the $l\neq 0$ case, the potential (1) can not be solved
exactly without an approximation to the centrifugal term [22]. Hence, in the
previous papers, several approximations have been developed to find better
analytical formulas for the hyperbolical potential [27].

Our aim in this work is to attempt to study the arbitrary $l$-state
solutions of the Schr\"{o}dinger equation for the hyperbolical potential. In
order to improve the accuracy of our previous approximation [20,21], we
propose and apply a new approximation scheme for the centrifugal term in the
form:%
\begin{equation}
\frac{1}{r^{2}}\approx 4\alpha ^{2}\left[ c_{\text{0}}+\frac{e^{-2\alpha r}}{%
1-e^{-2\alpha r}}+\left( \frac{e^{-2\alpha r}}{1-e^{-2\alpha r}}\right) ^{2}%
\right] ,
\end{equation}%
where $c_{0}$ is a proper shift to be found by the expansion procedures.
Thus, with this new approximation scheme, we calculate the $l\neq 0$ energy
levels and wavefunctions of the hyperbolical potential using the
hypergeometric approach (Nikiforov and Uvarov) N-U method. This method has
shown its power in calculating the exact energy levels for some solvable
quantum systems [13,14,20-22]. The approximation given by (2) has proved its
power and accuracy over the other currently used approximations in
literature [16-21]. It has been applied recently on the Manning-Rosen
potential [22] and has also proved its power and efficiency when compared
with the other numerical simulations for the non-approximated problem used
to calculate the energy bound states. It provides good results which are in
agreement with the numerical integration method by Lucha and Sch\"{o}berl
[28].

The paper is organized as follows: In Section II we breifly present the N-U
method. In Section III, we present the new proposed approximation scheme and
apply it to calculate the $l$-wave eigensolutions of the SE for the
hyperbolical potential by the N-U method. In Section IV, we present our
numerical results for energy eigenvalues numerically for arbitrary quantum
numbers $n$ and $l$ with two different values of the potential parameter $%
\sigma _{\text{0}}$. Section V, is devoted to study two special cases, the $%
s $-wave ($l=0)$ case and the $\sigma _{0}=1$ exponential-type potential.
Finally, we make a few concluding remarks in Section VI.

\section{The Nikiforv-Uvarov Method}

The Nikiforov-Uvarov (N-U) method is based on solving the second-order
linear differential equation by reducing it to a generalized equation of
hypergeometric type [29]. In this method after employing an appropriate
coordinate transformation $z=z(r),$ the Schr\"{o}dinger equation can be
written in the following form:%
\begin{equation}
\psi _{n}^{\prime \prime }(z)+\frac{\widetilde{\tau }(z)}{\sigma (z)}\psi
_{n}^{\prime }(z)+\frac{\widetilde{\sigma }(z)}{\sigma ^{2}(z)}\psi
_{n}(z)=0,
\end{equation}%
where $\sigma (z)$ and $\widetilde{\sigma }(z)$ are the polynomials with at
most of second-degree, and $\widetilde{\tau }(s)$ is a first-degree
polynomial. The special orthogonal polynomials [29] reduce Eq. (3) to a
simple form by employing $\psi _{n}(z)=\phi _{n}(z)y_{n}(z),$ and choosing
an appropriate function $\phi _{n}(z).$ Consequently, Eq. (3) can be reduced
into an equation of the following hypergeometric type:%
\begin{equation}
\sigma (z)y_{n}^{\prime \prime }(z)+\tau (z)y_{n}^{\prime }(z)+\lambda
y_{n}(z)=0,
\end{equation}%
where $\tau (z)=\widetilde{\tau }(z)+2\pi (z)$ (its derivative must be
negative) and $\lambda $ is a constant given in the form%
\begin{equation}
\lambda =\lambda _{n}=-n\tau ^{\prime }(z)-\frac{n\left( n-1\right) }{2}%
\sigma ^{\prime \prime }(z),\text{\ \ \ }n=0,1,2,...
\end{equation}%
It is worthwhile to note that $\lambda $ or $\lambda _{n}$ are obtained from
a particular solution of the form $y(z)=y_{n}(z)$ which is a polynomial of
degree $n.$ Further, $\ y_{n}(z)$ is the hypergeometric-type function whose
polynomial solutions are given by Rodrigues relation%
\begin{equation}
y_{n}(z)=\frac{B_{n}}{\rho (z)}\frac{d^{n}}{dz^{n}}\left[ \sigma ^{n}(z)\rho
(z)\right] ,
\end{equation}%
where $B_{n}$ is the normalization constant and the weight function $\rho
(z) $ must satisfy the differential equation: [29]%
\begin{equation}
w^{\prime }(z)-\left( \frac{\tau (z)}{\sigma (z)}\right) w(z)=0,\text{ }%
w(z)=\sigma (z)\rho (z).
\end{equation}%
In order to determine the weight function given in Eq. (7), we must obtain
the following polynomial:%
\begin{equation}
\pi (z)=\frac{\sigma ^{\prime }(z)-\widetilde{\tau }(z)}{2}\pm \sqrt{\left(
\frac{\sigma ^{\prime }(z)-\widetilde{\tau }(z)}{2}\right) ^{2}-\widetilde{%
\sigma }(z)+k\sigma (z)}.
\end{equation}%
In principle, the expression under the square root sign in Eq. (8) can be
arranged as the square of a polynomial. This is possible only if its
discriminant is zero. In this case, an equation for $k$ is obtained. After
solving this equation, the obtained values of $k$ are included in the N-U
method and here there is a relationship between $\lambda $ and $k$ by $%
k=\lambda -\pi ^{\prime }(z).$ After this point an appropriate $\phi _{n}(z)$
can be extracted from the differential equation:%
\begin{equation}
\phi ^{\prime }(z)-\left( \frac{\pi (z)}{\sigma (z)}\right) \phi (z)=0.
\end{equation}

\section{Analytical Solutions}

\subsection{An Impoved Approximation Scheme}

The approximation is based on the expansion of the centrifugal term in a
series of exponentials depending on the intermolecular distance $r$ and
keeping terms up to second order. Therefore, instead of using the
approximation in [9,11,19], we use this choice of approximation:
\begin{equation*}
\frac{1}{r^{2}}\approx (2\alpha )^{2}\left[ c_{0}+v(r)+v^{2}(r)\right] ,%
\text{ }v(r)=\frac{e^{-2\alpha r}}{1-e^{-2\alpha r}},
\end{equation*}%
\begin{equation}
\frac{1}{r^{2}}\approx (2\alpha )^{2}\left[ c_{0}+\frac{1}{e^{2\alpha r}-1}+%
\frac{1}{\left( e^{2\alpha r}-1\right) ^{2}}\right] ,
\end{equation}%
which has a similar form of the hyperbolical potential. Changing the
coordinate to $x$ by using $x=(r-r_{0})/r_{0},$ one obtains%
\begin{equation}
\left( 1+x\right) ^{-2}=\gamma ^{2}\left[ c_{0}+\frac{1}{e^{\gamma (1+x)}-1}+%
\frac{1}{\left( e^{\gamma (1+x)}-1\right) ^{2}}\right] ,\text{ }\gamma
=2\alpha r_{0}
\end{equation}%
and e$xpanding$ Eq. (11) around $r=r_{0}$ $(x=0),$ we obtain the following
Taylor's expansion:
\begin{equation*}
1-2x+O(x^{2})=\gamma ^{2}\left( c_{0}+\frac{1}{e^{\gamma }-1}+\frac{1}{%
\left( e^{\gamma }-1\right) ^{2}}\right)
\end{equation*}%
\begin{equation}
-\gamma ^{3}\left( \frac{1}{e^{\gamma }-1}+\frac{3}{\left( e^{\gamma
}-1\right) ^{2}}+\frac{2}{\left( e^{\gamma }-1\right) ^{3}}\right)
x+O(x^{2}),
\end{equation}%
from which we obtain
\begin{equation*}
\gamma ^{2}\left[ c_{0}+\frac{1}{e^{\gamma }-1}+\frac{1}{(e^{\gamma }-1)^{2}}%
\right] =1,
\end{equation*}%
\begin{equation}
\gamma ^{3}\left( \frac{1}{e^{\gamma }-1}+\frac{3}{\left( e^{\gamma
}-1\right) ^{2}}+\frac{2}{\left( e^{\gamma }-1\right) ^{3}}\right) =2.
\end{equation}%
Therefore the shifting paramete, $c_{0},$ can be found from the solution of
the above two equations as:
\begin{equation}
c_{0}=\frac{1}{\gamma ^{2}}-\frac{1}{e^{\gamma }-1}-\frac{1}{(e^{\gamma
}-1)^{2}}=0.0823058167837972,
\end{equation}%
where $e$ is the base of the natural logarithms, $e=2.718281828459045$ and
the parameter $\gamma =0.4990429999.$ Hence, we have the following
substitution for the centrifugal term:%
\begin{equation}
\frac{1}{r^{2}}=\underset{\alpha \rightarrow 0}{\lim }4\alpha ^{2}\left[
\frac{1}{\gamma ^{2}}-\frac{1}{e^{\gamma }-1}-\frac{1}{(e^{\gamma }-1)^{2}}+%
\frac{e^{-2\alpha r}}{1-e^{-2\alpha r}}+\left( \frac{e^{-2\alpha r}}{%
1-e^{-2\alpha r}}\right) ^{2}\right] .
\end{equation}%
Finally, it is worth to note that, in the case if $c_{0}=0,$ the
approximation given in Eq. (10) is identical to the commonly used
approximation in the previous works [9,11,19-21].

\subsection{Energy Eigenvalues and Eigenfunctions Solution}

To study any quantum physical system characterized by the empirical
potential given in Eq. (1), we solve the original $\mathrm{SE}$ which is
given in the well known textbooks [1,2]%
\begin{equation}
\left( \frac{p^{2}}{2m}+V(r)\right) \psi (\mathbf{r,}\theta ,\phi )=E\psi (%
\mathbf{r,}\theta ,\phi ),
\end{equation}%
where the potential $V(r)$ is taken as the hyperbolical potential (1). Using
the separation method with the wavefunction $\psi (\mathbf{r,}\theta ,\phi
)=r^{-1}R(r)Y_{lm}(\theta ,\phi ),$ we obtain the following radial Schr\"{o}%
dinger eqauation as%
\begin{equation*}
\frac{d^{2}R_{nl}(r)}{dr^{2}}+\left\{ \frac{2\mu E_{nl}}{\hbar ^{2}}-\frac{%
2\mu D}{\hbar ^{2}}\left[ 1-\sigma _{0}\left( \frac{e^{\alpha r}+e^{-\alpha
r}}{e^{\alpha r}-e^{-\alpha r}}\right) \right] ^{2}-4\alpha ^{2}l(l+1)\left(
c_{0}+\frac{e^{-2\alpha r}}{\left( 1-e^{-2\alpha r}\right) ^{2}}\right)
\right\}
\end{equation*}%
\begin{equation}
\times R_{nl}(r)=0.
\end{equation}%
Since the SE with the above hyperbolical potential has no analytical
solution for $l$-waves, we have used the approximation to the centrifugal
term given by case 1. The other approximations will be left for future
investigations. To solve it by the N-U method, we need to recast Eq. (17)
into the form of Eq. (3) changing the variables $r\rightarrow z$ through the
mapping function $r=f(z)$ and making the following definitions:%
\begin{equation}
z=e^{-2\alpha r},\text{ }\nu =\frac{\mu D}{2\alpha ^{2}\hbar ^{2}},\text{ }%
\varepsilon ^{\prime }=\sqrt{-\frac{\mu E_{nl}}{2\alpha ^{2}\hbar ^{2}}%
+\Delta E_{l}},\text{ }E_{nl}<\frac{2\alpha ^{2}\hbar ^{2}}{\mu }\Delta
E_{l},\text{ }\Delta E_{l}=l(l+1)c_{0},
\end{equation}%
we obtain the following hypergeometric equation:%
\begin{equation*}
\frac{d^{2}R(z)}{dz^{2}}+\frac{(1-z)}{z(1-z)}\frac{dR(z)}{dz}+\frac{1}{\left[
z(1-z)\right] ^{2}}
\end{equation*}%
\begin{equation}
\times \left\{ -\varepsilon ^{\prime }{}^{2}-\nu \left( 1-\sigma _{0}\right)
^{2}+\left[ 2\nu \left( 1-\sigma _{0}^{2}\right) +2\varepsilon ^{\prime
}{}^{2}-l(l+1)\right] z-\left[ \nu \left( 1+\sigma _{0}\right)
^{2}+\varepsilon ^{\prime }{}^{2}\right] z^{2}\right\} R(z)=0.
\end{equation}%
It is shown from Eq. (18) that for bound state (real) solutions, we require:%
\begin{equation}
z=\left\{
\begin{array}{ccc}
0, & \text{when} & r\rightarrow \infty , \\
1, & \text{when} & r\rightarrow 0,%
\end{array}%
\right.
\end{equation}%
and as a result the radial wavefunctions $R_{nl}(z)\rightarrow 0$ for the
values of $z$ given in Eq. (18). To apply the N-U method, we compare Eq.
(19) with Eq. (3) and obtain the following values for the parameters:
\begin{equation*}
\widetilde{\tau }(z)=1-z,\text{\ }\sigma (z)=z-z^{2},\text{\ }
\end{equation*}%
\begin{equation}
\widetilde{\sigma }(z)=-\left[ \nu \left( 1+\sigma _{0}\right)
^{2}+\varepsilon ^{\prime }{}^{2}\right] z^{2}+\left[ 2\nu \left( 1-\sigma
_{0}^{2}\right) +2\varepsilon ^{\prime }{}^{2}-l(l+1)\right] z-\varepsilon
^{\prime }{}^{2}-\nu \left( 1-\sigma _{0}\right) ^{2}.
\end{equation}%
If one inserts these values of parameters into Eq. (8), with $\sigma
^{\prime }(z)=1-2z,$ the following linear function is obtained%
\begin{equation}
\pi (z)=-\frac{z}{2}\pm \frac{1}{2}\sqrt{a_{2}z^{2}+a_{1}z+a_{0}},
\end{equation}%
where $a_{2}=1+4\left[ \varepsilon ^{\prime }{}^{2}+\nu \left( 1+\sigma
_{0}\right) ^{2}-k\right] ,$ $a_{1}=4\left\{ k+l(l+1)-2\nu \left( 1-\sigma
_{0}^{2}\right) -2\varepsilon ^{\prime }{}^{2}\right\} $ and $a_{0}=4\left[
\varepsilon ^{\prime }{}^{2}+\nu \left( 1-\sigma _{0}\right) ^{2}\right] .$
According to this method the expression in the square root has to be set
equal to zero, that is, $\Delta =a_{2}z^{2}+a_{1}z+a_{0}=0.$ Thus the
constant $k$ can be determined as%
\begin{equation}
k=-\frac{1}{4}\left[ (1+2l)^{2}+16\nu \sigma _{0}(\sigma _{0}-1)-1\right]
\pm \beta (1+2\delta ),
\end{equation}%
where%
\begin{equation}
\beta =\sqrt{\text{ }\varepsilon ^{\prime 2}+\nu \left( 1-\sigma _{0}\right)
^{2}},\text{ }\delta =\frac{1}{2}\left[ -1+\sqrt{16\nu \sigma
_{0}^{2}+(1+2l)^{2}}\right] .
\end{equation}%
In this regard, we can find four possible functions for $\pi (z)$ as%
\begin{equation*}
\pi (z)=-\frac{z}{2}\pm \frac{1}{2}
\end{equation*}%
\begin{equation}
\times \left\{
\begin{array}{c}
\left( 2\beta -2\delta -1\right) z-2\beta ,\text{ \ \ \ for \ \ }k=-\left[
(1+2l)^{2}+16\nu \sigma _{0}(\sigma _{0}-1)-1\right] /4+\beta (1+2\delta ),
\\
\left( 2\beta +2\delta +1\right) z-2\beta ;\text{ \ \ \ for \ \ }k=-\left[
(1+2l)^{2}+16\nu \sigma _{0}(\sigma _{0}-1)-1\right] /4-\beta (1+2\delta ).%
\end{array}%
\right.
\end{equation}%
We must select%
\begin{equation}
\text{\ }k=-\frac{z}{2}-\frac{1}{2}\left[ \left( 2\beta +2\delta +1\right)
z-2\beta \right] ,
\end{equation}%
in order to obtain the polynomial, $\tau (z)=\widetilde{\tau }(z)+2\pi (z)$
having negative derivative as%
\begin{equation}
\tau (z)=1-2z-\left[ 2\beta +2\delta +1\right] z-2\beta ,\text{ }\tau
^{\prime }(z)=-(2\beta +2\delta +3).
\end{equation}%
We can also write the values of $\lambda =k+\pi ^{\prime }(z)$ and $\lambda
_{n}=-n\tau ^{\prime }(z)-\frac{n\left( n-1\right) }{2}\sigma ^{\prime
\prime }(z),$\ $n=0,1,2,...$ as%
\begin{equation}
\lambda =-\frac{1}{4}\left[ (1+2l)^{2}+16\nu \sigma _{0}(\sigma _{0}-1)-1%
\right] -\left[ \beta +\delta +1\right] ,
\end{equation}%
\begin{equation}
\lambda _{n}=n(n+2\beta +2\delta +2),\text{ }n=0,1,2,\cdots
\end{equation}%
respectively. Additionally, using the definition of $\lambda =\lambda _{n}$
and solving the resulting equation for $\varepsilon ^{\prime },$ allows one
to obtain%
\begin{equation}
\beta =-\frac{(n+1)^{2}+l(l+1)+(2n+1)\delta -4\nu \sigma _{0}(1-\sigma _{0})%
}{2(n+\delta +1)}=\sqrt{-\frac{\mu E_{nl}}{2\alpha ^{2}\hbar ^{2}}+\Delta
E_{l}+\nu \left( 1-\sigma _{0}\right) ^{2}}.
\end{equation}%
Hence, we obtain analytically the following discrete bound-energy levels%
\begin{equation*}
E_{nl}=(1-\sigma _{0})^{2}D+\frac{2\alpha ^{2}\hbar ^{2}l(l+1)}{\mu }\left[
\frac{1}{\gamma ^{2}}-\frac{1}{e^{\gamma }-1}-\frac{1}{(e^{\gamma }-1)^{2}}%
\right]
\end{equation*}%
\begin{equation}
-\frac{2\alpha ^{2}\hbar ^{2}}{\mu }\left[ \frac{(n+1)^{2}+l(l+1)+(2n+1)%
\delta -4\nu \sigma _{0}(1-\sigma _{0})}{2(n+\delta +1)}\right] ^{2}\text{, }%
0\leq n,l<\infty
\end{equation}%
where $n=0,1,2,\cdots $ and $l$ signify the usual radial and angular
momentum quantum numbers, respectively.

Let us now find the corresponding radial part of the normalized wave
functions. Using $\sigma (z)$ and $\pi (z)$ in Eqs. (21) and (26), we obtain%
\begin{equation}
\phi (z)=z^{\beta }(1-z)^{\delta +1},
\end{equation}%
\begin{equation}
\rho (z)=z^{2\beta }(1-z)^{2\delta +1},
\end{equation}

\begin{equation}
y_{nl}(z)=C_{n}z^{-2\beta }(1-z)^{-(2\delta +1)}\frac{d^{n}}{dz^{n}}\left[
z^{n+2\beta }(1-z)^{n+2\delta +1}\right] .
\end{equation}%
The functions $\ y_{nl}(z)$, up to a numerical factor, are in the form of\
Jacobi polynomials, i.e., $\ y_{nl}(z)\simeq P_{n}^{(2\beta ,2\delta
+1)}(1-2z),$ valid physically in the interval $(0\leq r<\infty $ $%
\rightarrow $ $0\leq z\leq 1)$ [30]. Therefore, the radial part of the wave
functions can be found by substituting Eqs. (32) and (34) into $%
R_{nl}(z)=\phi (z)y_{nl}(z)$ as%
\begin{equation}
R_{nl}(r)=N_{nl}e^{-2\alpha \beta r}(1-e^{-2\alpha r})^{1+\delta
}P_{n}^{(2\beta ,2\delta +1)}(1-2e^{-2\alpha r}),
\end{equation}%
where $\beta $ and $\delta $ are given in Eq. (24) and $N_{nl}$ is a
normalization factor to be determined from the normalization condition:which
gives [20-22]%
\begin{equation*}
N_{nl}=\frac{1}{\sqrt{s(n)}},
\end{equation*}%
\begin{equation*}
s(n)=\frac{(-1)^{n}}{2\alpha }\frac{\Gamma (n+2\delta +2)\Gamma (n+2\beta
+1)^{2}}{\Gamma (n+2\beta +2\delta +2)}
\end{equation*}%
\begin{equation}
\times \sum\limits_{p,r=0}^{n}\frac{(-1)^{p+r}\Gamma (n+2\beta
+r-p+1)(p+2\delta +2)}{p!r!(n-p)!(n-r)!\Gamma (n+2\beta -p+1)\Gamma (2\beta
+r+1)(n+2\beta +r+2\delta +2)}.
\end{equation}

\section{Numerical Results}

To show the accuracy of the new approximation scheme, we calculate the
energy eigenvalues for various $n$ and $l$ quantum numbers with two
different values of the parameter $\sigma _{0}.$ The results calculated by
Eq. (31) are compared with those obtained by a MATHEMATICA package
programmed by Lucha and Sch\"{o}berl [28] as shown in Table 1 for
short-range potential (small $\alpha $) and long-range potential (large $%
\alpha $). It provides that the new proposed approximation scheme to the
centrifugal term in Eq. (2), even when the potential parameter $\alpha $
becomes large, produces energy eigenvalues of high accuracy like the other
numerical methods [28]. Consequently, this is also an illustration to assess
the validity and usefulness of our present approximation. Further, it is
quite simple, computationally efficient, reliable and accurate.

\section{Discussions}

We have used the hypergeometric method (N-U) to solve the radial $\mathrm{SE}
$ with the exponential-type potentials for arbitrary $l$-states$.$ We have
derived the binding energy spectra in Eq. (31) and their corresponding
normalized wave functions in Eq. (35).

Firstly, let us attempt to study the $s$-wave case $(l=0).$ Hence, the
energy eigenvalue and the radial eigen function solutions; (31) and (35),
reduce to the following forms:%
\begin{equation}
E_{n}=D(1-\sigma _{0})^{2}-\frac{2\alpha ^{2}\hbar ^{2}}{\mu }\left[ \frac{%
(n+1)^{2}+(2n+1)\delta _{1}-4\nu \sigma _{0}(1-\sigma _{0})}{2(n+\delta
_{1}+1)}\right] ^{2}\text{, }0\leq n<\infty ,
\end{equation}%
and%
\begin{equation}
R_{nl}(r)=N_{nl}e^{-2\alpha \beta _{1}r}(1-e^{-2\alpha r})^{1+\delta
_{1}}P_{n}^{(2\beta _{1},2\delta _{1}+1)}(1-2e^{-2\alpha r}),
\end{equation}%
respectively, with%
\begin{equation}
\beta _{1}=\sqrt{-\frac{\mu E_{n}}{2\alpha ^{2}\hbar ^{2}}+\frac{\mu D}{%
2\alpha ^{2}\hbar ^{2}}\left( 1-\sigma _{0}\right) ^{2}},\text{ }\delta _{1}=%
\frac{1}{2}\left[ -1+\sqrt{1+\frac{8\mu D\sigma _{0}^{2}}{\alpha ^{2}\hbar
^{2}}}\right] ,
\end{equation}%
and $N_{nl}$ is given in Eq. (36). This result is in agreement with Ref.
[27].

Secondly, we further discuss another special case $\sigma _{0}=1.$ As a
result, the potential (1) reduces to%
\begin{equation}
V(r)=\frac{4De^{-4\alpha r}}{\left( 1-e^{-2\alpha r}\right) ^{2}}.
\end{equation}%
The corresponding energy levels and radial wave functions are given by%
\begin{equation}
E_{nl}=\frac{2\alpha ^{2}\hbar ^{2}l(l+1)}{\mu }\left[ \frac{1}{\gamma ^{2}}-%
\frac{1}{e^{\gamma }-1}-\frac{1}{(e^{\gamma }-1)^{2}}\right] -\frac{2\alpha
^{2}\hbar ^{2}}{\mu }\left[ \frac{(n+1)^{2}+l(l+1)+(2n+1)\delta _{2}}{%
2(n+\delta _{2}+1)}\right] ^{2}\text{, }
\end{equation}%
and%
\begin{equation}
R_{nl}(r)=N_{nl}e^{-2\alpha \beta _{2}r}(1-e^{-2\alpha r})^{1+\delta
_{2}}P_{n}^{(2\beta _{2},2\delta _{2}+1)}(1-2e^{-2\alpha r}),
\end{equation}%
respectively, with%
\begin{equation}
\beta _{2}=\sqrt{-\frac{\mu E_{nl}}{2\alpha ^{2}\hbar ^{2}}+\frac{l(l+1)}{%
\gamma ^{2}}-\frac{l(l+1)}{e^{\gamma }-1}-\frac{l(l+1)}{(e^{\gamma }-1)^{2}}%
}{},\text{ }\delta _{2}=\frac{1}{2}\left[ -1+\sqrt{\frac{8\mu D}{\alpha
^{2}\hbar ^{2}}+(1+2l)^{2}}\right] ,
\end{equation}%
and $N_{nl}$ is given in Eq. (36). This result essentially coincides with
that of Manning-Rosen potential with the special case $A=0$ and $D=\frac{%
\alpha ^{3}(1-\alpha )\hbar ^{2}}{2\mu }$as shown in our recent work [22].

\section{Cocluding Remarks}

The arbirary $l$-wave solutions of the SE with an exponential-type potential
have been obtained approximately by proposing an improved shifted
approximation to the centrifugal term. It is found that the normalized wave
functions can be expressed by means of the Jacobi polynomials. With this
approximation scheme, we can easily build an analytic formulations [Eqs.
(31) and (35)] and use it to evaluate eigenvalues and eigenfunctions. This
approximation scheme has also been used with great success in problems which
do not have exact solutions for $l\neq 0$ case with exponential-type
potentials like Manning-Rosen potential and hyperbolic potential.
Essentially, two special cases for the $s$-wave case $(l=0)$ and $\sigma
_{0}=1$ and found that these results have been reduced to those given in
[25-27]. To show the accuracy of our results, we have calculated the
eigenvalues numerically for arbitrary $n$ and $l$ with two different values
of the parameter $\sigma _{0}.$ We found that the results obtained by (31)
are in good agreement with those obtained by using the MATHEMATICA program
based on the numerical integration procedure for short-range potential
(small $\alpha )$ and long-range potential (large $\alpha $) [28]. As a
demonstration of the accuracy of our results, Table 1 shows that the
estimated energy eigenvalues can be computed up to $0.001-0.13$ $\%.$
However, the accuracy of the energy eigenvalues in [27] is computed up to $%
0.051-1.0$ $\%.$ Therefore, the accuracy of the present model reaches up to $%
10-50$ times better than the estimations provided by [27].

\acknowledgments Work partially supported by the Scientific and
Technological Research Council of Turkey.

\newpage

{\normalsize %center
}

\newpage \baselineskip= 2\baselineskip% double space the text
%\end{document}
\bigskip \newpage

\begin{table}[tbp]
\caption{Energy eigenvalues as a function of the parameter $\protect\alpha $
for $2p,3p,3d,4p,4d,4f,5p,5d,5f,$ $5g,6p,6d,6f$ and $6g$ states for $\protect%
\sigma _{0}=0.1,$ $0.2$ and $D=10$ in atomic units $(\hbar =\protect\mu =1).$%
}%
\begin{tabular}{llllllll}
&  & $\sigma _{0}=0.1$ &  &  & $\sigma _{0}=0.2$ &  &  \\
states & $\alpha $ & present & Dong et al [27] & Lucha et al [28] & present
& Dong et al [27] & Lucha et al [28] \\
\tableline$2p$ & $0.10$ & $2.61874$ & $2.61556$ & $2.61935$ & $1.20876$ & $%
1.20559$ & $1.20903$ \\
& $0.15$ & $3.90544$ & $3.89830$ & $3.90645$ & $1.86636$ & $1.85922$ & $%
1.86689$ \\
& $0.20$ & $5.00331$ & $4.99062$ & $5.00457$ & $2.52000$ & $2.50731$ & $%
2.52080$ \\
& $0.25$ & $5.88594$ & $5.86611$ & $5.88725$ & $3.14666$ & $3.12683$ & $%
3.14766$ \\
$3p$ & $0.10$ & $4.73540$ & $4.73223$ & $4.73638$ & $2.68308$ & $2.67990$ & $%
2.68358$ \\
& $0.15$ & $6.04543$ & $6.03829$ & $6.04649$ & $3.67127$ & $3.66413$ & $%
3.67198$ \\
& $0.20$ & $6.91663$ & $6.90394$ & $6.91733$ & $4.46516$ & $4.45247$ & $%
4.46579$ \\
& $0.25$ & $7.48400$ & $7.46417$ & $7.48358$ & $5.09231$ & $5.07247$ & $%
5.09235$ \\
$3d$ & $0.10$ & $3.62699$ & $3.61747$ & $3.62769$ & $1.57873$ & $1.56921$ & $%
1.57920$ \\
& $0.15$ & $5.29404$ & $5.27263$ & $5.29510$ & $2.54773$ & $2.52631$ & $%
2.54859$ \\
& $0.20$ & $6.47492$ & $6.43684$ & $6.47598$ & $3.48119$ & $3.44311$ & $%
3.48228$ \\
$4p$ & $0.10$ & $6.00287$ & $5.99969$ & $6.00390$ & $3.75692$ & $3.75375$ & $%
3.75758$ \\
& $0.15$ & $7.11526$ & $7.10812$ & $7.11589$ & $4.81215$ & $4.80501$ & $%
4.81274$ \\
& $0.20$ & $7.71903$ & $7.70634$ & $7.71826$ & $5.53111$ & $5.51842$ & $%
5.53087$ \\
$4d$ & $0.10$ & $5.33129$ & $5.32177$ & $5.33216$ & $2.95257$ & $2.94305$ & $%
2.95317$ \\
& $0.15$ & $6.73583$ & $6.71441$ & $6.73642$ & $4.10410$ & $4.08268$ & $%
4.10470$ \\
& $0.20$ & $7.54480$ & $7.50672$ & $7.54331$ & $5.00179$ & $4.96371$ & $%
5.00137$ \\
$4f$ & $0.10$ & $4.68965$ & $4.67061$ & $4.69058$ & $2.07342$ & $2.05438$ & $%
2.07417$ \\
& $0.15$ & $6.42992$ & $6.38708$ & $6.43112$ & $3.35622$ & $3.31338$ & $%
3.35742$ \\
& $0.20$ & $7.43397$ & $7.35782$ & $7.43334$ & $4.47408$ & $4.39793$ & $%
4.47486$ \\
$5p$ & $0.10$ & $6.80345$ & $6.80027$ & $6.80432$ & $4.54946$ & $4.54628$ & $%
4.55015$ \\
$5d$ & $0.10$ & $6.37762$ & $6.36810$ & $6.37842$ & $3.95677$ & $3.94725$ & $%
3.95740$ \\
$5f$ & $0.10$ & $5.98063$ & $5.96159$ & $5.98147$ & $3.31497$ & $3.29593$ & $%
3.31567$ \\
$5g$ & $0.10$ & $5.62805$ & $5.59631$ & $5.62926$ & $2.64017$ & $2.60844$ & $%
2.64124$ \\
$6p$ & $0.10$ & $7.32416$ & $7.32099$ & $7.32476$ & $5.13763$ & $5.13446$ & $%
5.13824$ \\
$6d$ & $0.10$ & $7.04824$ & $7.03872$ & $7.04873$ & $4.69929$ & $4.68977$ & $%
4.69979$ \\
$6f$ & $0.10$ & $6.79479$ & $6.77575$ & $6.79528$ & $4.22654$ & $4.20751$ & $%
4.22706$ \\
$6g$ & $0.10$ & $6.57377$ & $6.54204$ & $6.57452$ & $3.73301$ & $3.70128$ & $%
3.73378$%
\end{tabular}%
\end{table}

\end{document}